# A novel minimal in vitro system for analyzing HIV-1 Gag mediated budding

Dong Gui[1], Sharad Gupta[1], Jun Xu[1], Roya Zandi[1], Sarjeet Gill[2], I-Chueh Huang[2], A.L.N. Rao[3], Umar Mohideen[1]

[1] Department of Physics & Astronomy, [2] Department of Cell Biology & Neuroscience, [3] Department of Plant Pathology & Microbiology, University of California, Riverside, CA, USA

Corresponding Author: Umar Mohideen, Department of Physics & Astronomy, University of California, Riverside, CA 92521. Tel. (951) 827 5390; email: umar.mohideen@ucr.edu

**ABSTRACT**

A biomimetic minimalist model membrane was used to study the mechanism and kinetics of cell-free *in vitro* HIV-1 Gag budding from a giant unilamellar vesicle (GUV). Real time interaction of Gag, RNA and lipid leading to the formation of mini-vesicles was measured using confocal microscopy. Gag forms resolution limited punctae on the GUV lipid membrane. Introduction of the Gag and urea to a GUV solution containing RNA led to the budding of mini-vesicles on the inside surface of the GUV. The GUV diameter showed a linear decrease in time due to bud formation. Both bud formation and decrease in GUV size were proportional to Gag concentration. In the absence of RNA, addition of urea to GUVs incubated with Gag also resulted in subvesicle formation but exterior to the surface. These observations suggest the possibility that clustering of GAG proteins leads to membrane invagination even in the absence of host cell proteins. The method presented here is promising, and allows for systematic study of the dynamics of assembly of immature HIV and help classify the hierarchy of factors that impact the Gag protein initiated assembly of retroviruses such as HIV.

**Keywords:** HIV-1 Gag, lipid Gag interaction, Gag multimerization, vesicle budding, vesicle formation rate

## Introduction

A remarkable feature of retroviruses such as the human immunodeficiency virus (HIV-I) is that, one structural polyprotein coded by the viral genome, Gag, together with the viral RNA lead to the spontaneous assembly of virus like particles in mammalian cells [1-6]. HIV-I Gag is a 55kD multi domain peptide with 4 primary structural domains which from the N to C terminus are: matrix (MA, P17), capsid (CA, P24), nucleo capsid (NC, P7) and p6, In mature particles the CA domain forms the shell around the virus, the NC domain consolidates and protects the RNA genome at the viral core. In the prevailing hypothesis [1,3,2], the multistep HIV-I assembly is initiated by viral RNA binding to the NC domain of Gag. The origin of the binding is thought to be negatively charged RNA attracted to positively charged MA and NC domains [7]. This complex is transported to the host plasma membrane [8] along with cellular proteins, such as the endosomal sorting complex required for transport (ESCRT) [9-11]. The Gag MA domain binds to the cytoplasmic plasma membrane through electrostatic forces resulting from a patch of basic residues [12] and insertion of hydrophobic covalently bound myristoyl fatty acid chain into the bilayer [13]. The binding is facilitated by negatively charged (anionic) PI(4,5)P2 lipids. The lipid and RNA binding causes Gag to unfold into a rod, which leads to tight hexagonal packing assembly [14,15,7,16,17]. Electron microscopy studies [18,1,16,14,19] show that Gag aggregates on the inner cytoplasmic leaflet of the plasma membrane lead to budding into the extracellular space and to the release of immature virus like particles. In these immature freshly budded spherical particles, the Gag proteins are radially located. Following the budding process, action of the viral protease leads to the cone shaped capsid of mature HIV-1.

Gag's central role in HIV replication and infectivity has been well established using *in vivo* and *in vitro* studies [1,6,3], but many questions remain open, particularly regarding the kinetics, reaction pathways and necessary cellular components involved. For example, mutant Gag, in which Gag-lipid binding, Gag-nucleic acid binding and Gag-Gag dimerization were independently modified in T-cells, showed that not all three interactions are needed but at least two were necessary for the formation of virus like particles [20]. This finding opens new questions on the combined role of Gag, nucleic acids and lipids in HIV formation.

A simple cell-free *in vitro* system would be highly beneficial in ascertaining the clear roles of the different Gag domains and their rates of interaction with lipids and RNA. This is mainly due to the fact that in *in vivo* studies, it is difficult to clearly interpret the key roles of RNA and lipids as Gag comes into contact with a number of cellular components that might rescue or alter the kinetics. For example, budding of virus like particles in HeLa cells has been investigated using Total Internal Reflection Fluorescence Microscopy (TIRFM) [21,22] and Atomic Force Microscopy (AFM) [23]. However, the cellular environment is difficult to control and it is frequently unclear whether the observed phenomena result from input Gag, RNA or due to the complex interaction of cellular factors, as numerous cellular proteins might interact with Gag during assembly [24]. Similarly many cell free systems, such as template directed assembly on gold nanoparticles [25,26], and wheat germs or reticulocyte lysates, which have been used to study retrovirus assembly, while useful, have the same disadvantages [27-30].

To investigate whether GAG induces membrane invaginations in the absence of RNA and cellular proteins and to analyze vesicle formation in a controlled manner we employed a minimal giant unilammellar vesicle (GUV) system. The mechanics and kinetics of Gag interaction with a lipid bilayer were monitored using confocal microscopy. In the absence of RNA and cellular proteins, Gag molecules aggregate and form mini-buds only in the presence of urea in solution, i.e., the sole interaction of Gag with membrane is not sufficient for formation of mini-vesicles. Further the addition of Gag along with yeast tRNA leads to multimerization of Gag and formation of Gag punctae, and budding of mini-vesicles followed by rapid disintegration of GUV into minivesicles, in the *absence* of any host cell proteins but with urea present. The system promises to be a very useful platform to study many of the critical factors, such as the role of the different Gag structural domains, lipids and RNA and to identify ligands and cofactors that effect and impact Gag induced assembly of HIV. Controls using unmyristoylated Gag P55, the capsid domain Gag P24, and urea, were done. We found that complete Gag P55 along with urea is necessary to observe budding in the GUVs. The essential role of urea in vesicle formation in the cell-free system suggests that conformational changes of the Gag protein mediated by other cellular factors maybe pivotal for virion egress.

**Materials & Methods**

**Materials**

The lipids DPhPC (1,2-diphytanoyl-sn-glycero-3-phosphocholine) and POPC (1-palmitoyl-2-oleoyl-sn-glycero-3-phosphocholine), cholesterol, and brain PI(4,5)P2 (L-α-phosphatidylinositol-4,5-bisphosphate) were from Avanti Polar Lipids (Alabaster, AL). For labeling membranes the fluorescent stain $DilC_{18}$ (1,1'-dioctadecyl-3,3,3',3'-tetramethylin dodicarbocyanine perchlorate) was from Invitrogen (Carlsbad, California). DyLight 488 NHS-Ester protein labeling kit and unmyristoylated Gag P55 was purchased from Thermo Scientific (Rockford, IL). Urea was from Fisher Scientific. The myristoylated Gag P55 reagent at 1 mg/ml concentration in PBS, was obtained through the NIH AIDS Reagent Program, Division of AIDS, NIAID, NIH: HIV-$1_{IIIB}$ pr55 Gag. The capsid domain HIV-1 p24 IIIB P24 at 1 mg/ml concentration in 20mM Na-Ac, 0.05M NaCl, pH 6.5 was also obtained through the NIH AIDS Reagent Program.
.
**Preparation of giant lipid vesicles:**

Giant vesicles were prepared by using the gentle hydration method. In short, a lipid mixture of DPhPC:POPC:PI(4,5)P$_2$:Cholesterol (molar ratio 11:4:1:4) was used and doped with 0.1 mol% of the lipid stain $DilC_{18}$. The mixture was dissolved in methanol/chloroform [2:1 (v/v)]. Lipids in the solvent were transferred to a clean roughed Teflon plate, which was placed in a beaker and sealed with aluminum foil. The solvent was then removed by placing the beaker in vacuum for 6 h. After that, argon gas saturated with water vapor at a temperature of 45 $^{o}$C (above the transition temperature of the lipid) was blown over the Teflon plate for 20 min and 10 ml of 200 mM sucrose in distilled water was added to the beaker. The sucrose was necessary for weight loading the GUV to keep it stable during confocal microscopy. The hydration process was allowed to occur overnight. Following hydration, "clouds" of vesicles were observed. The giant vesicles were harvested by aspirating with a Pasteur pipette and examined directly under a fluorescence microscope. These giant vesicles were stable for one week when stored at 4 $^{o}$C.

**Labeling of Gag with DyLight 488:**

Ten μl of borate buffer (0.67 M) was added to one 2.5 μl DyLight 488 vial. The solution was mixed with 25 μl of 1mg/ml myristoylated Gag P55 and vortexed gently for 1 min. The mixture was incubated in the dark for 1 h at room temperature. Excess DyLight was removed from the labeled Gag using centrifugation through a size exclusion column as suggested by the supplier. The same labelling procedure was used for unmyristoylated Gag P55 and the capsid domain Gag P24. Freshly labeled Gag was used in all the reported experiments. For experiments where myristoylated Gag with urea was added to the GUV solution, the 0.5 mg/ml labeled Gag was mixed with an equal volume of 12 M urea in PBS buffer and stored for 2 h. This mixture was then diluted by a factor of 5 with PBS buffer prior to use resulting in a 0.05 mg/ml myristoylated Gag P55 in a 1.2 M urea solution.

**Confocal Microscopy:**

Confocal microscopy was performed using a Leica SP5 microscope fitted with a 40X water immersion objective (N.A=1.1). Two excitation light sources with wavelengths of 488 nm and 633 nm were used. Collection light wavelengths filters at 450-550 nm in the green and 600-700 nm in the red were used in the results reported. The imaging plane was maintained at the equatorial plane of the vesicles. GUVs were added to an imaging cell containing two ml of 200 mM glucose in distilled water. The glucose balances the osmotic pressure of 200 mM sucrose inside the GUV. The heavier sucrose keeps the vesicle relatively immobile during imaging and the introduction of reactants. It was independently confirmed that the sucrose and glucose do not participate in any of the reactions.

**Controls:**

*Gag Capsid domain*: The effect of Gag P24, the capsid domain of Gag P55, on GUVs was studied. Here Gag P24 labelled with DyLight 488 was added to the GUV containing test cell. The GUVs were observed in the confocal microscope. The final concentration of Gag P24 in the test cell was 180 nM. Even at this concentration (10x of myristoylated Gag P55 used) no punctae on the GUVs were observed. The GUVs remained unchanged and no budding phenomena was detected. Thus the Gag P24 was observed to not bind to the lipid membrane of the GUV.

*Unmyristoylated Gag P55 in urea:* The necessity of N terminal myristoylation of Gag P55 for binding to lipid membranes and the formation of virus like particles has been well studied [31-33]. Unmyristoylated Gag is thus an excellent control to study Gag mediated vesicle budding in GUVs. The experiment reported below with myristoylated Gag P55 in urea was repeated with unmyristoylated Gag P55 and found to have no effects on GUVs. Here 0.05 mg/ml unmyristoylated Gag P55 in urea solution was added to the GUV test cell in steps till the concentration reached 28 nM Gag in 42 mM urea. The GUVs remained unaffected when observed under the confocal microscope. Even though the concentration of unmyristoylated Gag and urea reached nearly three times that of the case of myristoylated Gag no budding or decrease of the GUV surface area was observed.

*Urea*: The effect of only urea on the GUVs was also studied. 0.6 M urea in PBS was added to the GUV containing test cell in 10 µl steps till the concentration reached 15 mM (the same as used with myristoylated Gag P55). Again the GUVs remained unaltered and no vesicle budding on the GUV surface or decrease in the GUV diameter with time was observed. No change in GUVs were observed even when the urea concentration was increased to 20 mM. The same was repeated for yeast RNA containing GUVs. Again no change due to the addition of urea was observed.

## Results & Discussion

Unilamellar and multilamellar GUVs labeled with $DilC_{18}$ (red fluorescent lipid stain) were prepared as described in Material and Methods. HIV Gag p55 was labeled with DyLight 488 (green fluorescence). Confocal images were collected at excitation laser light (488 nm and 633nm) and with emission wavelength filters (450-550 nm (green) and 600-700 nm (red)).

**Gag interaction with GUV:**

We investigated the binding of myristoylated Gag P55 to the GUV lipid membrane. It has been shown [34,35] that the MA domain of Gag has a binding site for PI(4,5) P2 anionic lipid, and GAG specifically binds $PI(4,5)P_2$ containing GUVs [10]. Gag binds lipids through electrostatic and hydrophobic interactions. Basic residues of the MA domain, which electrostatically bind to anionic lipids and the N terminal hydrophobic myristate group is conjectured to insert into bilayers [31-33]. To verify that GAG specifically binds GUV lipids, we incubated DyLight 488 labeled myristoylated Gag (50nM) with $DilC_{18}$ labeled GUV solution. After an hour of incubation the GUVs were inspected with a confocal microscope. As observed in Fig. 1, midplane GUV images show the presence of bright green punctae on the bilayer membrane. No budding or local distortions of the GUV profile due to the presence of Gag could be observed. Gag punctae were relatively uniformly sized with a diameter of 900 nm, and punctae sizes were similar to that observed by Carlson and Hurley [10]. The observed sizes were not true sizes due to the resolution limit and image smearing from thermal and mechanical motion of the vesicle surface. This is confirmed from the fact that the observed GUV membrane thickness nominally of 5 nm was about 600-700 nm in the image. Note that the confocal microscope resolution limit is around 500 nm. The lack of punctae larger than 900 nm indicates that there was an optimum preferred size for Gag puncta on the lipid membrane. We believe that the Gag molecules aggregate after their insertion into the membrane. An alternate hypothesis is the insertion of fully formed Gag particulates into the GUV membrane. Insertion of whole Gag clusters into the GUV membrane is not as probable due to size. The uniformity of punctae observed and absence of larger aggregates on the lipid membranes points to their probable assembly in the bilayer. This maximum size of the Gag puncta could be the result of a competition between the spontaneous curvature of Gag proteins, the membrane bending rigidity and tension, and unfavorable energetic costs at the domain boundary. As a control, the effect of Gag P24, the capsid domain of Gag P55, on GUVs was studied. As above the Gag P24 labelled with DyLight 488 was added to the GUV containing test cell. Even at a 180 nM final concentration of Gag P24 in the test cell no punctae on the GUVs were observed.

**Dynamics of Gag puncta:**

When two puncta approached each other, they appeared to have a repulsive force between them. The force is probably of elastic origin resulting from the membrane deformation induced by bound Gag. This repulsion leads to the relatively uniform size distribution of the Gag punctae observed. Such elastic repulsive forces were observed in dimpled lipid domains on GUV [36], and result from competition between membrane bending stiffness and boundary line tension. Gag puncta were found to undergo Brownian motion on the vesicle surface. The translational diffusion of a particle due to Brownian motion on a planar fluid surface such as a lipid bilayer membrane is given by $<r^2> = 4\,Dt$, where $r$ is the distance traveled in time interval $t$ and $D$ is the diffusion coefficient [37]. Since the distance travelled at in our experiments was much smaller than the GUV radius, we ignored the curvature of space when measuring $D$. We calculated the mean squared displacement of Gag puncta $<r^2>$ along the vesicle surface as observed in confocal images for each time interval $t=0.75$ s. The diffusion coefficient $D$ calculated from averaging $r^2$ for five puncta over 80 intervals was found to be $0.17\pm0.02$ $\mu m^2$/sec. The error was the standard error of the mean. Embedded proteins on an artificial two-dimensional membrane have been observed to undergo Brownian motion [38]. The measured diffusion coefficient $D$, with the time and distance resolution as observed here, for single proteins varied from 4-0.58 $\mu m^2$/ sec in artificially constituted lipid membranes [38]. The observed $D$ for Gag puncta is 6-24 times less than that for monomers on bilayer membranes. It is known that in the case of two-dimensional lipid like fluids, the diffusion coefficient $D$ is inversely related to the particle diameter [37]. The observed small value of the diffusion coefficient in Gag is probably due to its multimerization on the lipid membrane. A decrease by a factor of 6-14 in the diffusion constant is consistent with a multimer size that is between 36-576 times that of the monomer. Thus the value of $D$ obtained is consistent with multimer formation in the presence of lipids as observed in NMR [39], circular dichroism (CD) [40], electron microscopy [41,42], small angle neutron scattering (SANS) [40] and neutron reflectivity [7].

**RNA interaction with GUV:**

Next we investigated the interaction of RNA on the GUV lipid membrane. GUVs of diameter around 30 μm were imaged in a confocal microscope using a two ml test cell. The equatorial plane of the GUV appeared as uniform circles with a resolution limited thickness of around 500 nm (Fig. 2a). Then 20 μl of 20 mg/ml yeast tRNA were added to the GUV solution resulting in a 5 μM RNA concentration. The vesicle was observed for 10 mins. A typical image is shown in Fig. 2b. No visible changes in the shape of the GUV were observed due to the presence of RNA. Some inhomogeneities in the fluorescence intensity along the vesicle surface were noted indicating that RNA-lipid interactions led to lipid density changes. The dense regions of around 1 μm in size were relatively immobile in comparison to the Gag puncta observed in Fig. 1. Their diffusion rate on the GUV surface was much lower with a value of $0.10\pm0.02$ $\mu m^2$/ sec, tracking four samples for 40 time intervals. No change in the GUV diameter with time was observed. The lipid density inhomogeneities might be due to the presence of lipid raft like micro domain platforms, which form the substrate for future vesicle budding. Such micro-domain formation has been hypothesized as a precursor to budding [43]. Further investigations are necessary to confirm the properties and origin of these inhomogeneities.

**Gag and RNA interaction with GUV:**

Next the role of Gag in vesicle budding with RNA containing GUV was studied using the same two ml sample cell. A solution with 0.05 mg/ml Gag in 1.2 M urea was added to the GUV test cell in 5 μl steps. After the addition of 25 μl corresponding to a 10 nM Gag concentration (with 15 mM urea) solution, modifications of the GUV surface could be observed. A typical modification is shown in Fig. 2c. In addition to the inhomogeneities in fluorescence observed with RNA in Fig. 2b, buds of diameter $2.5\pm0.5$ μm appeared on the inside surface of the GUV. One such bud is indicated with an arrow in Fig. 2c. The appearance of these buds corresponded to a decrease in the GUV diameter. The change in GUV diameter as a function of time is shown in Fig. 4. Over 240 s the rate of change of the diameter is linear. This will be discussed in more detail in the section below. Control experiments with just RNA and the same volume of 1.2 M urea introduced into the GUV solution led to no changes in the GUVs. Additional control experiments using a myristoylated Gag P55 solution without urea were done and found to have no effect on the RNA containing GUVs.

When the Gag concentration in the GUV solution was increased to 20 nM, a rapid change of the GUV profile was observed as shown in Fig. 3a-d. As illustrated in the figure, additional buds appeared on the inside surface. The resolution limited average diameter of the buds was 2.2 ±0.5 μm. Here the error corresponds to the instrument resolution as typified by the thickness of the vesicle bilayer membrane in the image. These buds underwent Brownian motion on the GUV surface, with a diffusion coefficient $D$ of 0.11±0.02 μm$^2$/s, for four samples tracked for 61 time intervals. The lower diffusion rates observed for the buds show that they are distinctly different in structure and membrane site location from the Gag puncta observed in Fig. 1. The buds tended to aggregate in one location on the GUV surface. The location of the aggregate is indicated with an arrow in Fig. 3c,d. The aggregate size increases in time as observed in Fig. 3c-d consistent with shrinking GUV diameter from 18 to 11 μm in 180 s. Also note the clear presence of the green labeled Gag on the buds and the aggregate. This observation that the buds stay attached to the GUV surface, *i.e.*, do not detach and escape from the GUV, is consistent with the absence of host cell proteins, including ESCRT, in the experiment. In HIV-1 replication, the ESCRT complex of the host is engaged by the p6 domain of the Gag affecting the scission of the buds from the plasma membrane [44-46,43,47,48].

The bud formation on the inside surface of the GUV after the introduction of the Gag is consistent with the evolution of mini-vesicles. In immature HIV-1 the budding process is thought to be energetically driven by the self assembly of the CA domains of Gag into hexamers [16,14]. The rigid rod like shape of the Gag induces membrane curvature. Membrane interaction with myristoylated MA domain provides the mechanical force driving membrane shape changes. Since the size resolution of confocal microscopy is only 0.5 μm, one cannot say with certainty that these mini-vesicles correspond to virus like particles. All these observations indicate are that Gag proteins introduced into the solution outside the GUV bind to the outside lipid leaflet of the GUV. Gag multimerization then leads to membrane invagination towards the GUV center, resulting in the observed buds on the inside surface. Based on lipid volume conservation of the original 33 μm diameter GUV, the aggregate has to be composed of hollow mini-vesicles, as the amount of solid lipid forming the initial GUV is much less than the 10x10x5 μm$^3$ aggregate size that is observed.

It is instructive to compare the similar budding phenomena that have been observed in other systems such as Shiga toxin (STxB) [49] and ESCRT [43,50]. In the case of STxB, the invagination of tubules by the glycolipid binding B-subunit of bacterial STxB in GUV composed of DOPC, cholesterol and porcine Gb3 glycolipid was investigated [49]. The spontaneous clustering of the toxin on the GUV surface was imaged by confocal microscopy. The clusters of STxB resulted in tubular invaginations on lowering of the membrane tension due to the GUV surface increase from Bodipy$_{FL}$-C$_5$-HPC photo activation or by increasing the osmotic pressure of the bath [49]. The authors proposed the STxB-Gb3 cluster complex led to a negative curvature [49] similar to the case of aggregation of Gags observed in this paper. In the case of host cell proteins, such as ESCRT, interacting with GUV, the models project formation of microdomain clusters leading to an increase in the line tension along the domain boundaries resulting in buckling and formation of a bud when its exceeds a threshold level. The physical mechanism that leads to membrane budding with ESCRTs is presently not understood. The ESCRT complexes observed on the neck of the bud form a scaffold with an optimum size necessary for scission. These steps predict a large energy cost of 100 $K_BT$ necessary for the complete release of the minivesicles [50]. In the studies here, as host cell proteins are absent, the large energy costs for scission of the buds from the vesicle prevent their dissociation from the surface of the GUV.

**Gag interaction and rate kinetics with RNA containing GUV:**
The change in the GUV diameter as a function of time is shown in Fig. 4. Time t=0 relates to the addition of Gag corresponding to a 10 nM Gag concentration into the GUV test cell. The solid line is a smooth fit to the data and meant as an aid to the eye. The change in the diameter is close to linear over the entire period of 240 s the GUV was studied at this Gag concentration. The rate of change in the GUV diameter is -0.035 µm/s. Next the Gag concentration in the test cell was increased to 20 nM after 240 s, as indicated by the arrow in Fig. 4. The diameter undergoes a rapid change, probably due to the nonequilibrium Gag concentration caused by the injection of Gag. As the Gag solution was pipetted in the vicinity of the GUV, the concentration was higher than 20 nM immediately after its introduction. After equilibration, determined to be around 90 s after Gag addition, the rate of decrease in the GUV diameter in the linear region was found to be -0.067 µm/s. Thus an increase in the Gag concentration from 10 nM to 20 nM resulted in a rate of change of GUV diameter from 0.035 to 0.067 µm/s. A higher Gag concentration is expected to promote multimerization [51] leading to increased budding. The rate increase in vesicle formation along with the decrease of the diameter as a function of Gag concentration is consistent with enhanced vesicle budding.

**Effect of urea and rate kinetics with Gag containing GUV:**
We sought to clarify the role of urea in the Gag solution used to understand the interaction with RNA containing GUV above. Towards this end we incubated DyLight 488 labeled myristylated Gag (50nM) with DilC$_{18}$ labeled GUV solution for an hour, and the GUVs were analyzed with a confocal microscope. Similar to Fig. 1, images of the GUV had uniform bright green Gag punctae on the bilayer membrane as shown in Fig. 5a and no budding or local distortions of the GUV surface could be observed. Next five µl of 6 M urea were added to the two ml test cell (15 mM urea final concentration) with the Gag containing GUVs. The GUVs rapidly shrunk and disintegrated into mini-vesicles within 10 s of addition of the urea. The sequence of time events for one such GUV is shown in Fig. 5. In contrast to the case with the presence of RNA (Fig. 3), no buds were observed on the interior wall of the GUV. Only subvesicles were observed attached to the exterior GUV wall after urea addition as shown in Fig. 5c. Urea is a well known chaotropic agent used for the solubilization of proteins. Its role in the activation of Gag bound to the GUV membrane might be due to conformational changes introduced by the urea interaction. The corresponding change in the GUV diameter as a function of time is shown in Fig. 6. The rate of change of diameter was 0.6 µm/s. This is 20 times faster than that observed with RNA containing GUVs.

As mentioned in Methods and Materials, various controls were done. In particular, unmyristoylated Gag P55 and the capsid domain Gag P24 were prepared in the same manner as the myristoylated Gag P55 with urea above and the experiments repeated. The GUVs remained unaltered and no budding phenomena was observed. N terminal myristoylation of Gag P55 has been shown necessary for binding to lipid membranes and formation of virus like particles [31-33]. Thus the absence of myristoyl group in the controls prevents their insertion into the GUV membrane. The action of urea of the same concentration as used above on the GUVs was also studied and found to have no effect.

The rates of HIV-1 assembly and release in HeLa cells have been studied with Total Internal Reflection Fluorescence Microscopy (TIRFM) [22,21]. Atomic force microscopy studies of retroviral budding were done on murine leukemia virus (MLV) [23]. In TIRFM the fluorescence saturation time for Gag-GFP puncta was used as a marker and found to be 5-9 mins (exponential time constant found to be 233 s [22]). The time taken for a rapid increase in velocity of these fluorescence units which was attributed to the virion release was found to be around 25 min [22]. Note that only the formation of the Gag puncta could be detected by TIRFM. The budding of the virus like particles can only be inferred. The budding time scales for MLV in mouse fibroblast cells were found to be around 45 mins [23]. These values are not directly comparable to the experiments reported on the GUV, as Gag and viral RNA have to be expressed and transported across the cells. Diffusion rates alone in the cytoplasm are factor of 10 smaller in cells and thus the virus assembly rates would be correspondingly less. Thus the time scales for budding observed in GUVs are probably consistent with those observed *in vivo* after accounting for the diffusion rates.

**Conclusion**

We used a highly simplified minimalist experimental system of a GUV for the cell-free *in vitro* study of HIV-1 particle assembly mechanism and kinetics. Real time interaction of Gag, RNA and lipid was measured using confocal microscopy. Gag was found to lead to resolution limited punctae on the GUV lipid membranes. Even though, due to the optical resolution limit, the curvature of the domains cannot be observed, based on the physical arguments, we conclude that Gag aggregates form domains that are curved. This is based on the fact that flat domains would diffuse and coalesce to form even larger aggregates. However, in this system Gag punctae repel each other and their sizes remain finite. The size of a puncta depends on the spontaneous curvature of Gag proteins, the membrane bending rigidity and tension, and unfavorable energetic costs at the domain boundary.

We also found that RNA interacting with GUV caused lipid density inhomogeneities on the GUV surface, which might be precursors to raft formation. The introduction of Gag in urea to a GUV solution containing RNA led to the budding of mini-vesicles on the inside surface of the GUV. The GUV diameter decreased due to bud formation. The corresponding rate of decrease of the GUV diameter was found to be linear in time. The bud formation and the decrease in GUV size were found to be proportional to the Gag concentration. In the absence of RNA, the addition of urea to GUV membranes incubated with Gag also led to the formation of subvesicles. Controls using unmyristoylated Gag P55 with urea, the capsid domain Gag P24 and urea alone were found not to alter the GUVs. In all cases, the GUV remained stable in the absence of myristoylated Gag P55 and urea. The overall approach used here will allow for a systematic study of the dynamics and help classify the hierarchy of factors that impact the Gag protein initiated assembly of retroviruses. In future studies, we will explore the role of the different domains of Gag and their interplay with lipids and RNA using mutant Gag.

**Acknowledgment:** The authors acknowledge grant support from UCLABs Fund (D.G., S.G., J.X., U.M.), from the National Science Foundation through Grant No. DMR-13-10687 (RZ) and the UCR Chancellors Strategic Fund and Collaborative Research Seed Grant (D.G., U.M., R.Z., S.G, I-C.H, A.L.N. R).

**References:**

1. Rein, A., Datta, S.A.K., Jones, C.P., Musier-Forsyth, K.: Diverse interactions of retroviral Gag proteins with RNAs. Trends in Biochemical Sciences **36**(7), 373-380 (2011).
2. Chukkapalli, V., Ono, A.: Molecular determinants that regulate plasma membrane association of HIV-1 Gag. Journal of Molecular Biology **410**(4), 512-524 (2011).
3. Jouvenet, N., Simon, S.M., Bieniasz, P.D.: Visualizing HIV-1 Assembly. Journal of Molecular Biology **410**(4), 501-511 (2011). doi:http://dx.doi.org/10.1016/j.jmb.2011.04.062
4. Swanstrom, R., Wills, J.W.: Synthesis, assembly, and processing of viral proteins. Retroviruses (Coffin, J. M., Hughes, S. H. and Varmus, H. E., eds), 72 (1997).
5. Ganser-Pornillos, B.K., Yeager, M., Sundquist, W.I.: The structural biology of HIV assembly. Current Opinion in Structural Biology **18**(2), 203-217 (2008).
6. Briggs, J.A., Kräusslich, H.-G.: The molecular architecture of HIV. Journal of molecular biology **410**(4), 491-500 (2011).
7. Datta, S.A.K., Heinrich, F., Raghunandan, S., Krueger, S., Curtis, J.E., Rein, A., Nanda, H.: HIV-1 Gag extension: Conformational changes require simultaneous interaction with membrane and nucleic acid. Journal of Molecular Biology **406**(2), 205-214 (2011).
8. Gelderblom, H.R., Bauer, P.G., Özel, M., Höglund, P., Niedrig, M., Renz, H., Morath, B., Lundquist, P., Nilsson, Å., Mattow, J., Grund, C., Pauli, G.: Membrane Interactions of HIV. Wiley-Liss, New York (1992)
9. Henne, W.M., Buchkovich, N.J., Emr, S.D.: The ESCRT pathway. Dev Cell **21**(1), 77-91 (2011).
10. Carlson, L.A., Hurley, J.H.: In vitro reconstitution of the ordered assembly of the endosomal sorting complex required for transport at membrane-bound HIV-1 Gag clusters. Proceedings of the National Academy of Sciences **109**(42), 16928-16933 (2012). doi:10.1073/pnas.1211759109
11. Wollert, T., Wunder, C., Lippincott-Schwartz, J., Hurley, J.H.: Membrane scission by the ESCRT-III complex. Nature **458**(7235), 172-177 (2009).
12. Hill, C.P., Worthylake, D., Bancroft, D.P., Christensen, A.M., Sundquist, W.I.: Crystal structures of the trimeric human immunodeficiency virus type 1 matrix protein: implications for membrane association and assembly. Proceedings of the National Academy of Sciences **93**(7), 3099-3104 (1996).
13. Ono, A., Freed, E.O.: Binding of human immunodeficiency virus type 1 Gag to membrane: role of the matrix amino terminus. Journal of Virology **73**(5), 4136-4144 (1999).
14. Wright, E.R., Schooler, J.B., Ding, H.J., Kieffer, C., Fillmore, C., Sundquist, W.I., Jensen, G.J.: Electron cryotomography of immature HIV-1 virions reveals the structure of the CA and SP1 Gag shells. EMBO Journal **26**(8), 2218-2226 (2007).
15. Fuller, S.D., Wilk, T., Gowen, B.E., Kräusslich, H.-G., Vogt, V.M.: Cryo-electron microscopy reveals ordered domains in the immature HIV-1 particle. Current Biology **7**(10), 729-738 (1997).
16. Briggs, J.A.G., Riches, J.D., Glass, B., Bartonova, V., Zanetti, G., Kräusslich, H.G.: Structure and assembly of immature HIV. Proceedings of the National Academy of Sciences (2009). doi:10.1073/pnas.0903535106
17. Spearman, P., Horton, R., Ratner, L., Kuli-Zade, I.: Membrane binding of human immunodeficiency virus type 1 matrix protein in vivo supports a conformational myristyl switch mechanism. Journal of Virology **71**(9), 6582-6592 (1997).
18. Campbell, S., Rein, A.: In vitro assembly properties of human immunodeficiency virus type 1 Gag protein lacking the p6 domain. Journal of Virology **73**(3), 2270-2279 (1999).

19. Benjamin, J., Ganser-Pornillos, B.K., Tivol, W.F., Sundquist, W.I., Jensen, G.J.: Three-dimensional structure of HIV-1 virus-like particles by electron cryotomography. Journal of molecular biology **346**(2), 577-588 (2005).
20. O'Carroll, I.P., Soheilian, F., Kamata, A., Nagashima, K., Rein, A.: Elements in HIV-1 Gag contributing to virus particle assembly. Virus Research **171**(2), 341-345 (2013).
21. Jouvenet, N., Bieniasz, P.D., Simon, S.M.: Imaging the biogenesis of individual HIV-1 virions in live cells. Nature **454**(7201), 236-240 (2008).
22. Ivanchenko, S., Godinez, W.J., Lampe, M., Kräusslich, H.-G., Eils, R., Rohr, K., Bräuchle, C., Müller, B., Lamb, D.C.: Dynamics of HIV-1 assembly and release. PLoS Pathog **5**(11), e1000652 (2009).
23. Gladnikoff, M., Rousso, I.: Directly monitoring individual retrovirus budding events using atomic force microscopy. Biophysical Journal **94**(1), 320-326 (2008).
24. Ott, D.E.: Cellular proteins in HIV virions. Reviews in Medical Virology **7**(3), 167-180 (1997).
25. Vieweger, M., Goicochea, N., Koh, E.S., Dragnea, B.: Photothermal Imaging and Measurement of Protein Shell Stoichiometry of Single HIV-1 Gag Virus-like Nanoparticles. ACS nano **5**(9), 7324-7333 (2011).
26. Goicochea, N.L., Datta, S.A., Ayaluru, M., Kao, C., Rein, A., Dragnea, B.: Structure and stoichiometry of template-directed recombinant HIV-1 Gag particles. Journal of molecular biology **410**(4), 667-680 (2011).
27. Lingappa, J.R., Hill, R.L., Wong, M.L., Hegde, R.S.: A multistep, ATP-dependent pathway for assembly of human immunodeficiency virus capsids in a cell-free system. Journal of Cell Biology **136**(3), 567-581 (1997). doi:10.1083/jcb.136.3.567
28. Sakalian, M., Parker, S.D., Weldon, R.A., Hunter, E.: Synthesis and assembly of retrovirus Gag precursors into immature capsids in vitro. Journal of Virology **70**(6), 3706-3715 (1996).
29. Spearman, P., Ratner, L.: Human immunodeficiency virus type 1 capsid formation in reticulocyte lysates. Journal of Virology **70**(11), 8187-8194 (1996).
30. Weldon, R.A., Parker, W.B., Sakalian, M., Hunter, E.: Type d retrovirus capsid assembly and release are active events requiring ATP. Journal of Virology **72**(4), 3098-3106 (1998).
31. Bryant, M., Ratner, L.: Myristoylation-dependent replication and assembly of human immunodeficiency virus 1. Proceedings of the National Academy of Sciences **87**(2), 523-527 (1990).
32. Copeland, N.G., Jenkins, N.A., Nexø, B., Schultz, A.M., Rein, A., Mikkelsen, T., Jørgensen, P.: Poorly expressed endogenous ecotropic provirus of DBA/2 mice encodes a mutant Pr65gag protein that is not myristylated. Journal of Virology **62**(2), 479-487 (1988).
33. Rein, A., McClure, M.R., Rice, N.R., Luftig, R.B., Schultz, A.M.: Myristylation site in Pr65gag is essential for virus particle formation by Moloney murine leukemia virus. Proceedings of the National Academy of Sciences **83**(19), 7246-7250 (1986).
34. Saad, J.S., Miller, J., Tai, J., Kim, A., Ghanam, R.H., Summers, M.F.: Structural basis for targeting HIV-1 Gag proteins to the plasma membrane for virus assembly. Proceedings of the National Academy of Sciences **103**(30), 11364-11369 (2006). doi:10.1073/pnas.0602818103
35. Ono, A., Ablan, S.D., Lockett, S.J., Nagashima, K., Freed, E.O.: Phosphatidylinositol (4,5) bisphosphate regulates HIV-1 Gag targeting to the plasma membrane. Proceedings of the National Academy of Sciences **101**(41), 14889-14894 (2004). doi:10.1073/pnas.0405596101
36. Ursell, T.S., Klug, W.S., Phillips, R.: Morphology and interaction between lipid domains. Proceedings of the National Academy of Sciences **106**(32), 13301-13306 (2009).
37. Saffman, P.G., Delbrück, M.: Brownian motion in biological membranes. Proceedings of the National Academy of Sciences **72**(8), 3111-3113 (1975).
38. Kusumi, A., Nakada, C., Ritchie, K., Murase, K., Suzuki, K., Murakoshi, H., Kasai, R.S., Kondo, J., Fujiwara, T.: Paradigm shift of the plasma membrane concept from the two-dimensional continuum fluid to the partitioned fluid: High-speed single-molecule tracking of membrane molecules. Annual Review of Biophysics and Biomolecular Structure **34**(1), 351-378 (2005). doi:doi:10.1146/annurev.biophys.34.040204.144637
39. Amarasinghe, G.K., De Guzman, R.N., Turner, R.B., Chancellor, K.J., Wu, Z.R., Summers, M.F.: NMR structure of the HIV-1 nucleocapsid protein bound to stem-loop SL2 of the psi-RNA packaging signal. Implications for genome recognition. J Mol Biol **301**(2), 491-511 (2000).
40. Datta, S.A., Curtis, J.E., Ratcliff, W., Clark, P.K., Crist, R.M., Lebowitz, J., Krueger, S., Rein, A.: Conformation of the HIV-1 Gag protein in solution. J Mol Biol **365**(3), 812-824 (2007).
41. de Marco, A., Müller, B., Glass, B., Riches, J.D., Kräusslich, H.-G., Briggs, J.A.G.: Structural Analysis of HIV-1 Maturation Using Cryo-Electron Tomography. PLoS Pathog **6**(11), e1001215 (2010). doi:10.1371/journal.ppat.1001215
42. Carlson, L.-A., de Marco, A., Oberwinkler, H., Habermann, A., Briggs, J.A.G., Kräusslich, H.-G., Grünewald, K.: Cryo Electron Tomography of Native HIV-1 Budding Sites. PLoS Pathog **6**(11), e1001173 (2010). doi:10.1371/journal.ppat.1001173
43. Wollert, T., Hurley, J.H.: Molecular mechanism of multivesicular body biogenesis by ESCRT complexes. Nature **464**(7290), 864-869 (2010).
44. Huang, M., Orenstein, J.M., Martin, M.A., Freed, E.O.: p6Gag is required for particle production from full-length human immunodeficiency virus type 1 molecular clones expressing protease. Journal of Virology **69**(11), 6810-6818 (1995).
45. Bieniasz, P.D.: Late budding domains and host proteins in enveloped virus release. Virology **344**(1), 55-63 (2006).
46. Chen, B.J., Lamb, R.A.: Mechanisms for enveloped virus budding: Can some viruses do without an ESCRT? Virology **372**(2), 221-232 (2008).
47. Teis, D., Saksena, S., Emr, S.D.: SnapShot: The ESCRT Machinery. Cell **137**(1), 182-182.e181 (2009).
48. Ku, P.-I., Bendjennat, M., Ballew, J., Landesman, M.B., Saffarian, S.: ALIX is recruited temporarily into HIV-1 budding sites at the end of Gag assembly. PLoS ONE **9**(5), e96950 (2014).
49. Römer, W., Berland, L., Chambon, V., Gaus, K., Windschiegl, B., Tenza, D., Aly, M.R.E., Fraisier, V., Florent, J.-C., Perrais, D., Lamaze, C., Raposo, G., Steinem, C., Sens, P., Bassereau, P., Johannes, L.: Shiga toxin induces tubular membrane invaginations for its uptake into cells. Nature **450**(7170), 670-675 (2007).
50. Różycki, B., Boura, E., Hurley, J.H., Hummer, G.: Membrane-elasticity model of coatless vesicle budding induced by ESCRT complexes. PLoS Computational Biology **8**(10), e1002736 (2012).
51. Perez-Caballero, D., Hatziioannou, T., Martin-Serrano, J., Bieniasz, P.D.: Human immunodeficiency virus type 1 matrix inhibits and confers cooperativity on Gag precursor-membrane interactions. Journal of Cell Biology **78**(17), 9560-9563 (2004). doi:10.1128/jvi.78.17.9560-9563.2004

**Figure Captions**

Figure 1: Green labeled HIV-I myristoylated Gag P55 punctae can be observed on PI(4,5) $P_2$ containing GUV lipid membranes. The lipid membrane of the GUV is stained with 0.1 mol % of red fluorophore DilC$_{18}$. Gag P55 was incubated with the GUV for 1 h prior to imaging.

Figure 2: Action of RNA on the GUV lipid membrane. The midplane of a typical GUV (a) before and (b) after addition of yeast tRNA. The inhomogeneities in the fluorescence intensity caused by the addition of RNA probably correspond to density modifications in the bilayer lipid membrane. (c) Image after addition of myristoylated Gag P55 to a 10 nM concentration to GUV in (b). Vesicle budding (arrow pointing to one) on the inside surface of the GUV can be observed.

Figure 3: Action of Yeast tRNA and green labeled myristoylated Gag P55 on GUVs leading to mini vesicle formation and shrinkage. The images (a)-(d) show the same GUV as a function of time in one minute intervals after Gag concentration reaches 20 nM in the test cell. In (a) vesicle budding with green labeled Gag can be observed on the inside surface of the GUV. From (a)-(d) the diameter of the GUV shrinks due to the budding. The budded vesicles appear to aggregate in one location shown by arrow.

Figure 4: Interaction of myristoylated Gag P55 with RNA containing GUV. The same GUV was used for both concentrations. The GUV diameter decreased as a function of time due to the budding of mini-vesicles on the inside surface. After equilibration, the rate of decrease in the GUV diameter is linear in time and Gag concentration. Thus an increase in the Gag concentration from 10 nM to 20 nM resulted in a rate of change of GUV diameter from 0.035 to 0.067 µm/s. The solid line is a smooth fit to the data to be used as an aid to the eye.

Figure 5: Change in GUV on addition of five µl of 6 M urea to the two ml test cell with myristoylated Gag P55 containing GUVs. The GUVs rapidly shrunk and disintegrated into mini-vesicles within 10 s of addition of the urea.

Figure 6: Interaction of urea with myristoylated Gag P55 containing GUV. The GUV diameter decreased as a function of time due to the formation of sub-vesicles exterior to the surface of the GUV. After equilibration, the GUV diameter decreases almost linearly in time at the rate of 0.6 µm/s. The solid line is a smooth fit to the data to be used as an aid to the eye.

**Figure 1**

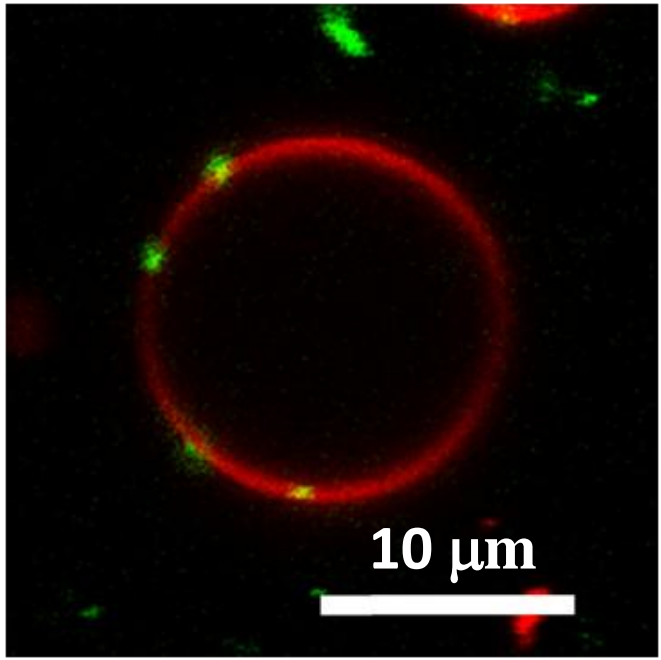

**Figure 2**

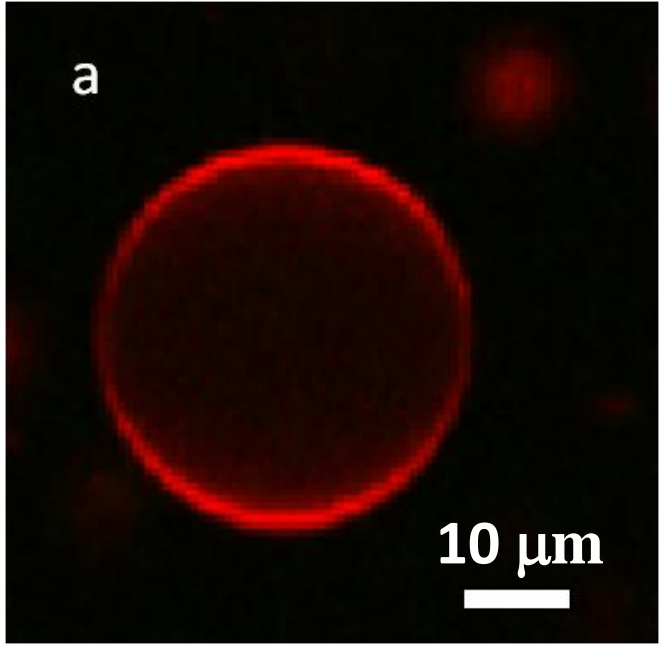

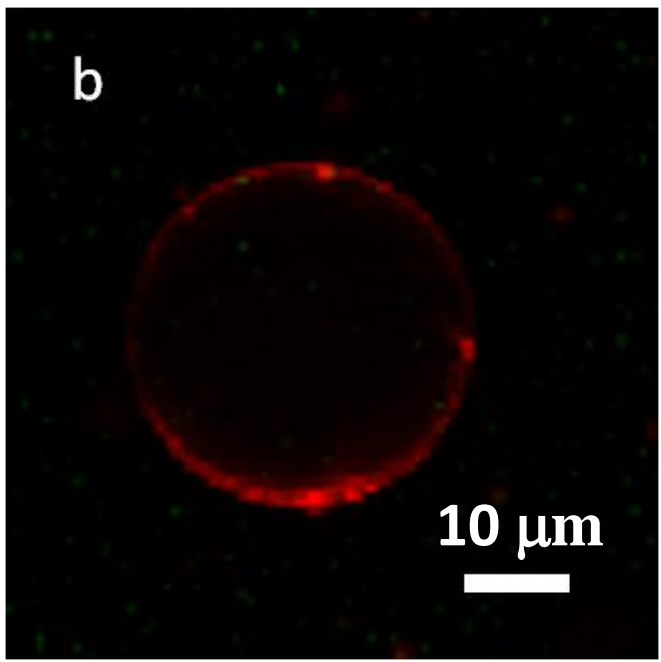

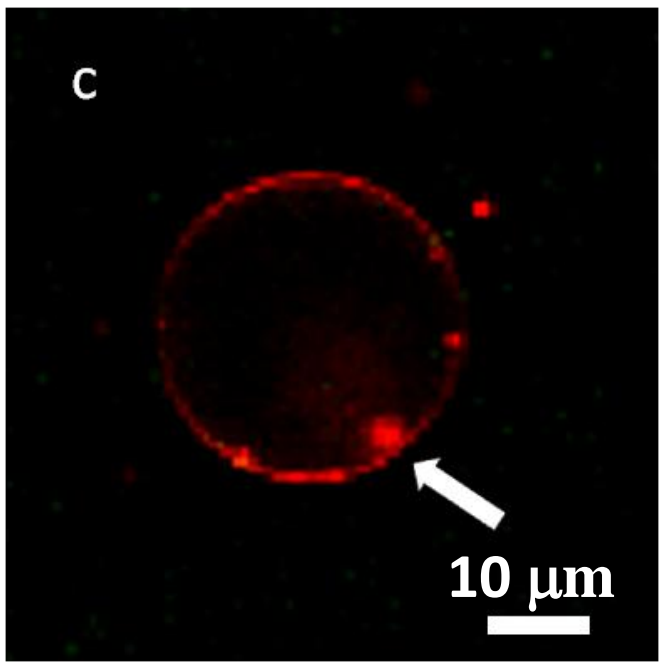

**Figure 3**

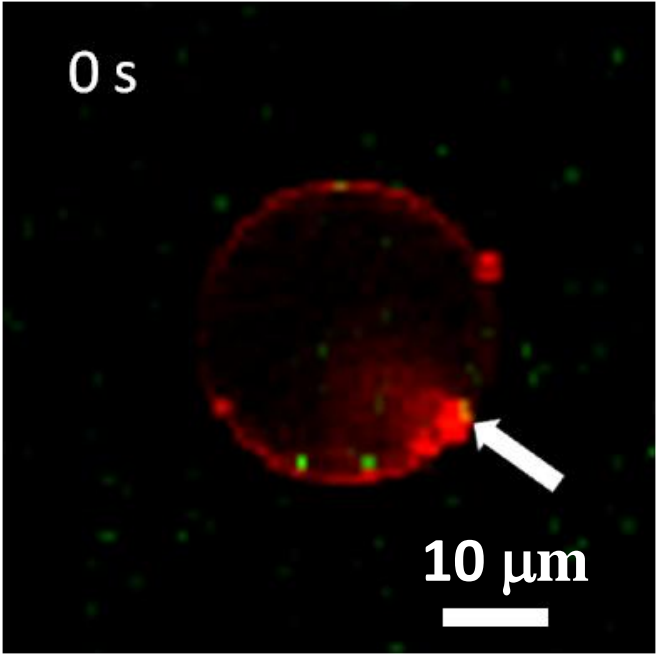

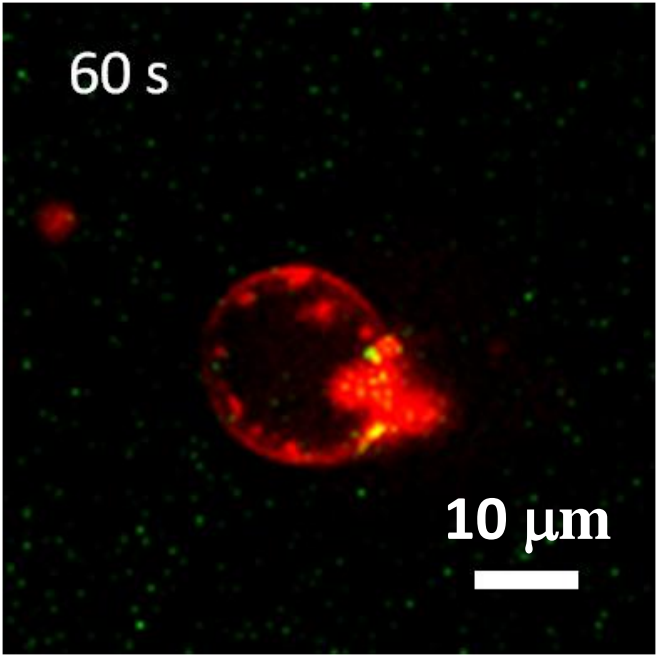

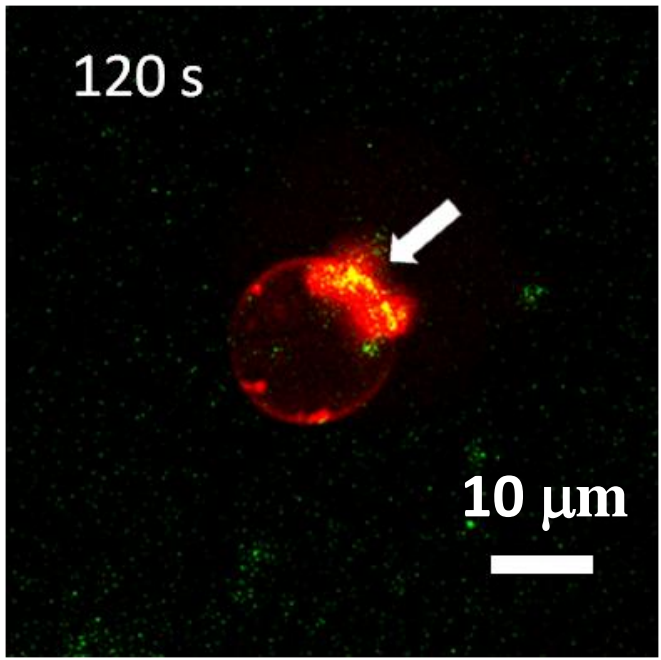

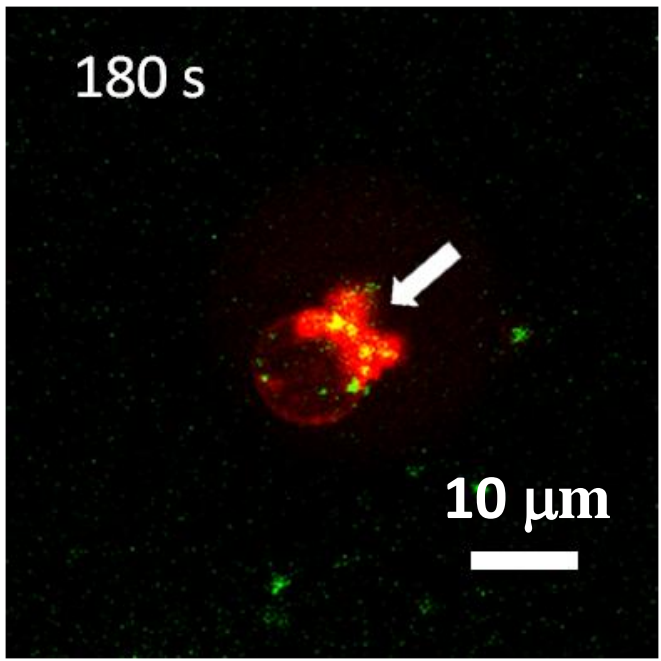

**Figure 4**

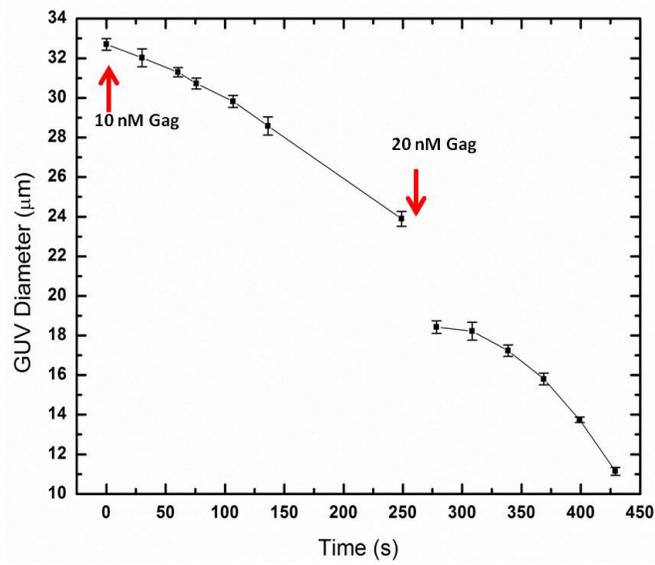

**Figure 5**

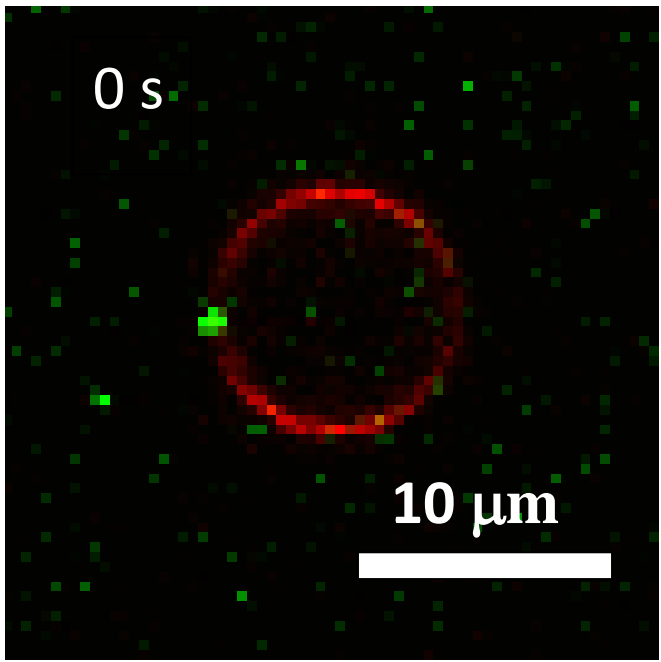

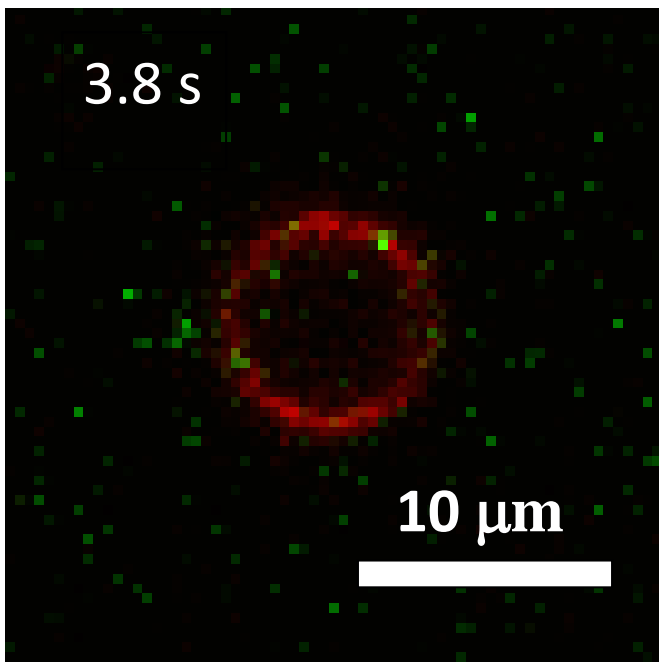

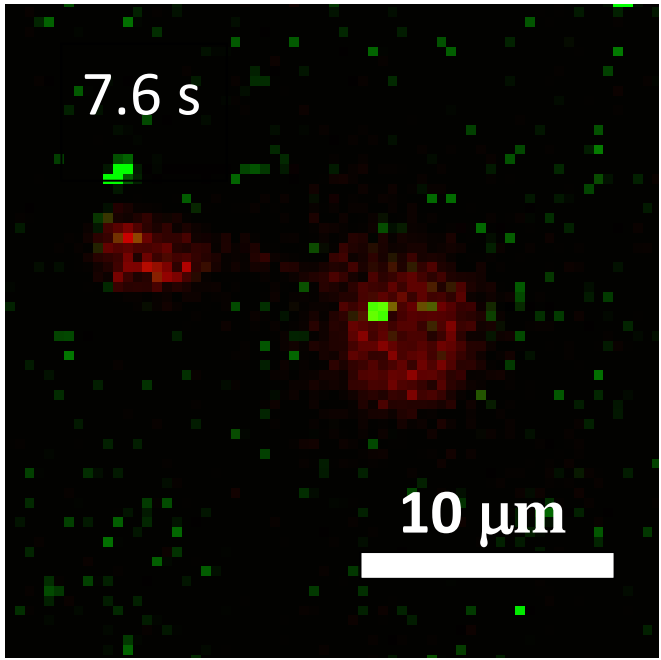

**Figure 6**

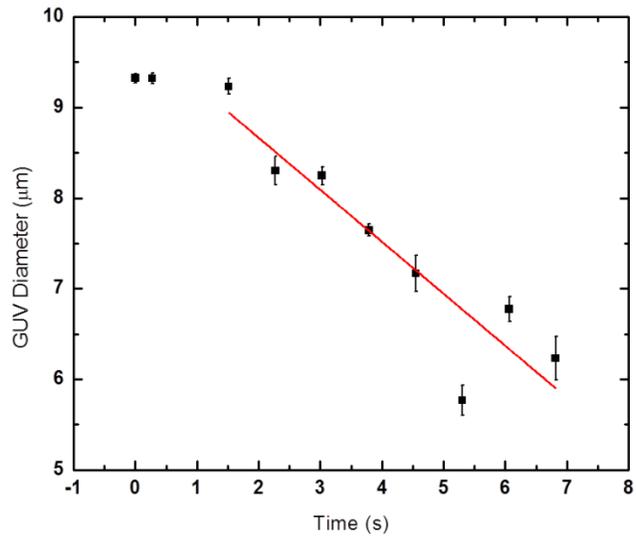

# A novel minimal in vitro system for analyzing HIV-1 Gag mediated budding

Dong Gui[1], Sharad Gupta[1], Jun Xu[1], Roya Zandi[1], Sarjeet Gill[2], I-Chueh Huang[2], A.L.N. Rao[3], Umar Mohideen[1]

[1] Department of Physics & Astronomy, [2] Department of Cell Biology & Neuroscience, [3] Department of Plant Pathology & Microbiology, University of California, Riverside, CA, USA

Corresponding Author: Umar Mohideen, Department of Physics & Astronomy, University of California, Riverside, CA 92521. Tel. (951) 827 5390; email: umar.mohideen@ucr.edu

Results from controls performed are provided below. We studied the effect of: (i) Gag P24 the capsid domain of HIV-1 Gag P55 on GUVs, (ii) unmyristoylated Gag P55 in urea on GUVs and (iii) urea on GUVs. All three did not affect the GUV and no bud formation was observed under the confocal microscope. In the case of Gag P24 and unmyristoylated Gag P55, which were both labelled green, no interaction with the lipid membrane of the GUV was observed.

## Effect of Gag P24 the Capsid Domain of Gag P55

The effect of Gag P24, the capsid domain of Gag P55, on GUVs was studied. Here Gag P24 labelled with DyLight 488 in the same manner as myristoylated Gag P55 was added to the GUV containing test cell. The GUVs were observed in the confocal microscope (Fig. S1(a)) prior to the addition of Gag P24. Next, 10 µl of the labeled Gag P24 were added to the test cell till the final concentration was 180 nM. As shown in Fig. S1(b), no changes in the GUVs were observed 41 minutes after the addition, even at this concentration, which is 10x of the concentration of myristoylated Gag P55 used. No punctae on the GUV lipid membrane were observed. Thus the Gag P24 does not bind the lipid membrane of the GUVs.

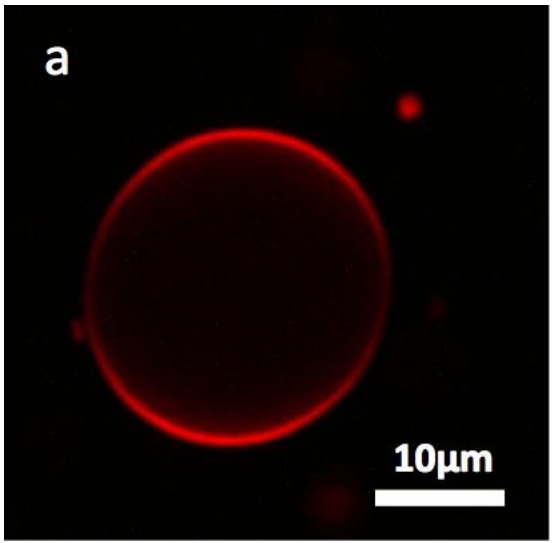
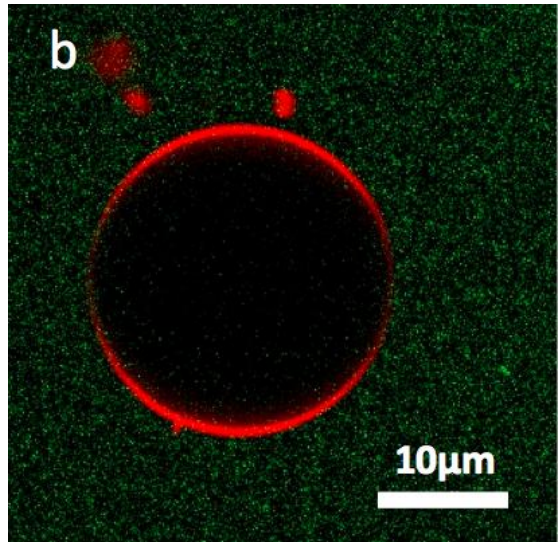

Fig. S1 (a)  Fig. S1 (b)

Figure S1: Effect of Gag P24, the capsid domain of Gag P55, on the GUV lipid membrane. Gag P24 is labeled green and the lipid membrane is labeled red. The midplane of a typical GUV (a) before and (b) 41 minutes after addition of green labeled Gag P24 in PBS. Gag P24 is observed to not bind the GUV lipid membrane and also has no effect on its structure.

# Effect of Unmyristoylated Gag P55 in urea on GUVs

We studied the necessity of N terminal myristoylation of Gag P55 for binding to lipid membranes and the formation of virus like particles. Unmyristoylated Gag is an excellent control to study the Gag mediated vesicle budding in GUVs. The same steps as in the experiment reported with myristoylated Gag P55 in urea was repeated with unmyristoylated Gag P55. Here green labelled 0.05 mg/ml unmyristoylated Gag P55 in urea was added in 5 µl steps first and then 10 µl steps to the test cell containing GUV till the final concentration reached 28 nM Gag in 42 mM urea. The GUVs remained unchanged even 19 minutes after the addition when observed under the confocal microscope. Note that with myristoylated Gag P55 in urea the vesicle is completely transformed within a minute. The observations are shown in Fig. S2. Even though the concentration of unmyristoylated Gag and urea reached nearly three times that of the case of myristoylated Gag no budding or decrease of the GUV surface area was observed.

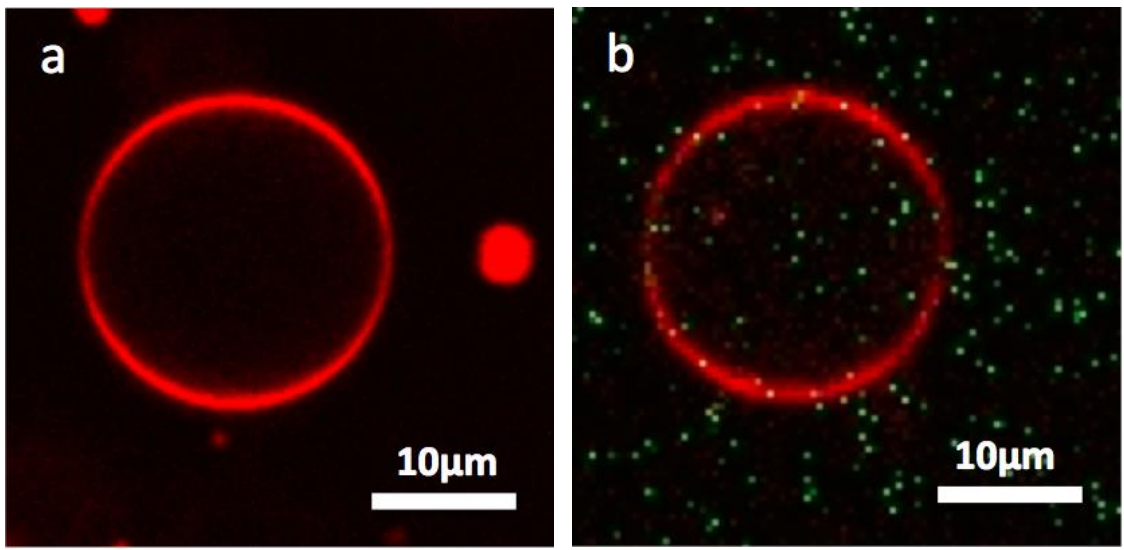

Fig. S2 (a)              Fig. S2 (b)

Figure S2: Effect of green labelled unmyristoylated Gag P55 in urea on the GUV lipid membrane. The midplane image of a typical GUV (a) before and (b) 19 mins after addition of green labelled unmyristoylated Gag P55 in urea. The final concentration reached 28 nM Gag P55 in 42 mM urea. No change in the GUVs is observed even though the concentration of Gag in urea is 3x times that of myristoylated Gag P55 in urea.

## Effect of Urea on GUVs

The effect of only urea on the GUVs was also studied. 0.6 M urea in PBS was added to the GUV containing test cell in 10 μl steps till the concentration reached 15 mM (the same as used with myristoylated Gag P55), while they were observed under the confocal microscope. As shown in Fig. S3, again the GUVs remained unaltered and no vesicle budding on the GUV surface or decrease in the GUV diameter with time was observed. No change in GUVs were observed even when the urea concentration was increased to 20 mM and observation time increased to 27 minutes. The same was repeated for yeast RNA containing GUVs. Again no change due to the addition of urea was observed.

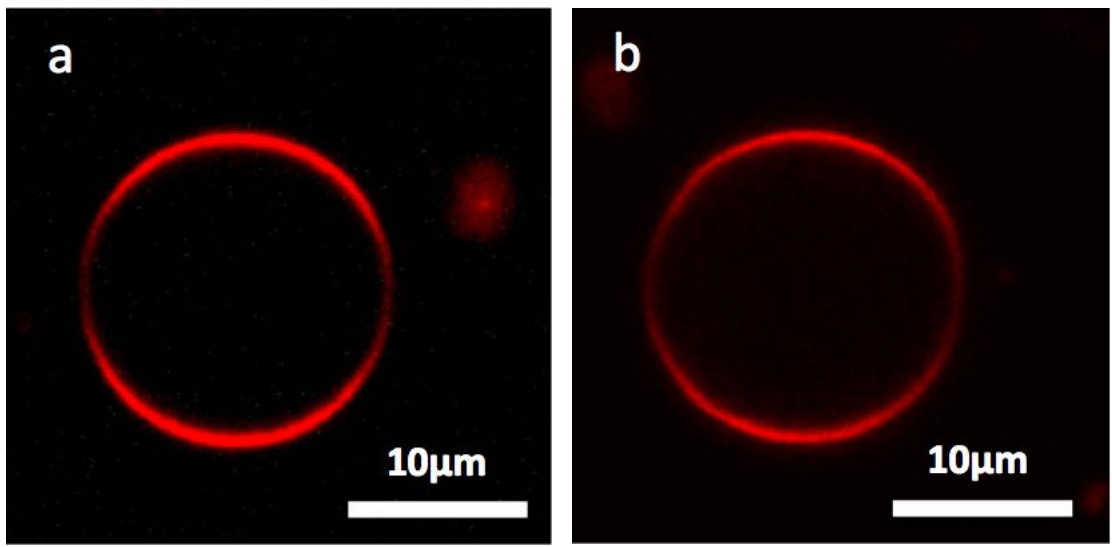

Fig. S3 (a)                                   Fig. S3 (b)

Figure S3: Effect of addition of urea to test cell containing GUVs. The midplane of a typical GUV under the confocal microscopy (a) before and (b) 27 minutes after addition of urea in PBS to a final concentration of 20 mM urea. No change in the GUVs is observed with the addition of urea.